\documentclass[12pt]{article}

\usepackage{graphicx}
\usepackage[a4paper,lmargin=1in,rmargin=1in,top=0.71in,bottom=0.71in]{geometry} 
\usepackage{setspace} 
\usepackage{amsmath}
\usepackage{amssymb}
\usepackage{bm} 
\usepackage{epsfig}
\usepackage{epstopdf} 
\usepackage{combelow} 
\usepackage{multirow} 
\usepackage{braket} 
\usepackage[
singlelinecheck=false 
]{caption} 

\usepackage{indentfirst} 
\usepackage[title]{appendix} 
\usepackage{titlesec} 
\titlelabel{\thetitle.\quad}
\titleformat*{\section}{\normalsize\bfseries}
\titleformat*{\subsection}{\normalsize\it}
\titleformat*{\subsubsection}{\normalsize\it}

\def\feq{\ensuremath{f^{(\mathrm{eq})}}}
\def\hmu{{\hat{\mu}}}
\def\hnu{{\hat{\nu}}}
\def\hlambda{{\hat{\lambda}}}
\def\hatt{{\hat{t}}}
\def\hati{{\hat{\imath}}}
\def\hatj{{\hat{\jmath}}}
\def\ti{{\tilde{\imath}}}
\def\tj{{\tilde{\jmath}}}

\begin{document}

\begin{center}
{\bf\uppercase
{Relativistic rotating Boltzmann gas using the tetrad formalism}
}
\end{center}

\begin{center}
{

Victor E.~Ambru\cb{s}${}^{\rm a, *}$, Robert Blaga${}^{\rm a, \dagger}$

\

{\footnotesize
{\it  ${}^{\rm a}$ Department of Physics, West University of Timi\cb{s}oara\\
  Bd.~Vasile P\^{a}rvan 4, Timi\cb{s}oara 300223, Romania}

  \

${}^{*}$ victor.ambrus@e-uvt.ro (corresponding author), ${}^\dagger$ robert.blaga@e-uvt.ro
}
}
\end{center}

\

\

\begin{center}
{
\footnotesize
\begin{tabular}{p{.30\linewidth}p{.60\linewidth}}
  {\bf Article info} &
  \multicolumn{1}{c}{\bf Abstract} \\
  {\it Received}:

  {\it Accepted}: 

  \


  {\bf PACS}:
  05.20.Dd, %
  47.75.+f. 
  &
  We consider an application of the tetrad formalism introduced by Cardall et al.~[Phys.~Rev.~D {\bf 88} (2013) 023011]
  to the problem of a rigidly rotating relativistic gas in thermal equilibrium and discuss the possible applications
  of this formalism to relativistic lattice Boltzmann simulations. We present in detail the transformation
  to the comoving frame, the choice of tetrad, as well as the explicit calculation and analysis of the components of the
  equilibrium particle flow four-vector and of the equilibrium stress-energy tensor.
 \end{tabular}
}
\end{center}

\

\onehalfspacing

\section{Introduction}\label{sec:intro}

In general relativity, the time component of the momentum $4$-vector of a particle of mass $m$ is
constrained to obey the mass-shell condition\footnote{Throughout this
paper, we use units in which $c = \hbar = K_B = 1$. We also use the mostly-plus convention for the metric signature.}:
\begin{equation}
 g_{\mu\nu} p^\mu p^\nu = -m^2.\label{eq:mshell}
\end{equation}
In relativistic kinetic theory, the spatial components $p^i$ are regarded as independent variables,
while $p^t$ is considered to depend on $p^i$ and the space-time metric components $g_{\mu\nu}$
through Eq.~\eqref{eq:mshell} \cite{cercignani02}.

When numerical algorithms are employed, this dependence of the form of $p^t$ on the space-time metric can
induce numerical complications. We refer in particular to the lattice Boltzmann method, where
the momentum space is discretised to provide an efficient quadrature tool for the evaluation of $N^\mu$
and $T^{\mu\nu}$ \cite{mendoza10prl}.
To avoid the cumbersome dependence of the mass-shell condition \eqref{eq:mshell} on the space-time characteristics,
non-holonomic tetrad fields $e_{\hmu}^\mu$ and the corresponding dual one-forms $\omega^{\hmu}_\mu$ can be introduced,
such that the resulting metric is the Minkowski metric $\eta_{\hmu\hnu} = {\rm diag}(-1, 1, 1, 1)$
\cite{lindquist66,riffert86,cardall15}:
\begin{equation}
 g_{\mu\nu} dx^\mu dx^\nu = \eta_{\hmu\hnu} \omega^\hmu \omega^\hnu.
\end{equation}
With respect to such a tetrad, the mass-shell condition \eqref{eq:mshell} reduces to the Minkowski
point-independent form:
\begin{equation}
 \eta_{\hmu\hnu} p^\hmu p^\hnu = -m^2.
\end{equation}
Moreover, the formalism of Ref.~\cite{cardall15} allows the spatial part of the momentum space to be
parametrised with respect to arbitrary coordinates, including spherical coordinates, which were employed
also for lattice Boltzmann simulations in Ref.~\cite{romatschke11}.

In this paper, we present an application of the tetrad formalism to the problem of a uniformly rotating relativistic
gas on Minkowski space. The aim of this paper is to illustrate the use of the tetrad formalism introduced
in Ref.~\cite{cardall15} to a simple problem, which we hope will facilitate its implementation in
future applications of the lattice Boltzmann method.

The paper is organised as follows. In section~\ref{sec:rboltz}, we briefly present the relativistic Boltzmann
equation and the properties of thermodynamic equilibrium for the cases of Maxwell-J\"uttner (M-J), Bose-Eintein (B-E)
and Fermi-Dirac (F-D) distributions. Section~\ref{sec:rot} presents the problem of a relativistic rotating gas,
solved using the tetrad formalism \cite{cardall15}. We also present a detailed analysis of the resulting
profiles of the particle number density $n$, energy density $\epsilon$ and pressure $P$ for the gas at
equilibrium when M-J, B-E and F-D statistics are employed. Section~\ref{sec:conc} concludes this paper.

\section{Relativistic Boltzmann equation}\label{sec:rboltz}

The Boltzmann equation for a relativistic gas can be written as \cite{cercignani02}:
\begin{equation}
 p^\mu \frac{\partial f}{\partial x^\mu} - \Gamma^i_{\mu\nu} p^\mu p^\nu \frac{\partial f}{\partial p^i} = C[f],
 \label{eq:boltz}
\end{equation}
where the one-particle distribution function $f \equiv f(x^\mu, p^i)$ is defined in terms of the space-time
coordinates $x^\mu$ and particle momentum components $p^\mu$ such that
\begin{equation}
 dN = f(x^\mu, p^i) \frac{p^t}{p_t} (-g) d^3x d^3p
\end{equation}
gives the total number of particles crossing the hypersurface element $d^3x = dx^1 dx^2 dx^3$ (of normal $dx^0$)
centred on $(x^1, x^2, x^3)$ at constant $x^0$, having momenta in a range
$dp^1 dp^2 dp^3$ about $(p^1, p^2, p^3)$ \cite{cercignani02}. In the above,
$g$ represents the determinant of the space-time metric $g_{\mu\nu}$.
The time component $p^t$ of the momentum
$4$-vector is fixed by the mass-shell condition \eqref{eq:mshell}.
The connection coefficients $\Gamma^i_{\mu\nu}$ appearing in Eq.~\eqref{eq:boltz}
have the following expression with respect to a holonomic frame:
\begin{equation}
 \Gamma^\lambda{}_{\mu\nu} = \frac{1}{2} g^{\lambda\sigma}\left(
 g_{\sigma\mu,\nu} + g_{\sigma\nu,\mu} - g_{\mu\nu,\sigma}\right),
\end{equation}
where a comma denotes differentiation with respect to the coordinates,
e.g.~$g_{\sigma\mu,\nu} \equiv \partial_{\nu} g_{\sigma\mu} \equiv
\frac{\partial g_{\sigma\mu}}{{\partial x^\nu}}$.

At local equilibrium, the collision integral $C[f]$ vanishes and
$f$ is given by \cite{cercignani02,romatschke12}:
\begin{equation}
 \feq \equiv \feq(Z; \beta, u_\mu) = Z \left[\exp\left( -\beta p^\mu u_\mu\right) - \varepsilon\right]^{-1},
 \label{eq:feq}
\end{equation}
where $Z$ represents the degrees of freedom, $\beta = 1 / T$ is the inverse local
temperature, $u_\mu$ are the covariant components of the macroscopic velocity
$4$-vector, and we have considered a vanishing chemical potential. The constant
$\varepsilon$ takes the values $-1$, $0$ and $1$  for the F-D, M-J and B-E distributions,
respectively. The conditions imposed on $\beta$ and $u^\mu$ such that $f = \feq$ can
be found by substituting Eq.~\eqref{eq:feq} into Eq.~\eqref{eq:boltz}:
\begin{equation}
 p^\mu \frac{\partial z}{\partial x^\mu} - \Gamma^i{}_{\mu\nu} p^\mu p^\nu
 \frac{\partial z}{\partial p^i} = 0,\label{eq:boltz_zeta}
\end{equation}
where
\begin{equation}
 z =  \beta p^\mu u_\mu.\label{eq:zdef}
\end{equation}
Keeping in mind that $p^t$ is a function
of $x^\mu$ and $p^i$, the mass-shell condition \eqref{eq:mshell} can be used to obtain
the following identities:
\begin{equation}
 \frac{\partial p^t}{\partial p^i} = - \frac{p_i}{p_t}, \qquad
 \frac{\partial p^t}{\partial x^\lambda} = -\frac{1}{p_t} \Gamma_{\mu\nu\lambda} p^\mu p^\nu.
 \label{eq:killing}
\end{equation}
Thus, Eq.~\eqref{eq:boltz_zeta} reduces to \cite{cercignani02}:
\begin{equation}
\frac{1}{2} p^\mu p^\nu \left[(\beta u_\mu)_{;\nu} +
 (\beta u_\nu)_{;\mu}\right] = 0.
\end{equation}
Since this relation must be valid for any $p^\mu$ we must have that $\beta u_\mu$ is a Killing vector
field of the space-time.

For a given particle distribution function $f$, the particle four-flow $N^\mu$ and the
stress-energy tensor $T^{\mu\nu}$ are given by\footnote{The factors of $(2\pi)^{-3}$ appear because
we follow Ref.~\cite{cardall15} and choose units in which $\hbar = 1$ instead of $h = 1$.}:
\begin{subequations}\label{eq:hydro}
\begin{align}
 N^\mu =& \frac{1}{(2\pi)^3} \int \frac{d^3p}{-p_t} \sqrt{-g}\, f\, p^\mu, \label{eq:Nmu}\\
 T^{\mu\nu} =& \frac{1}{(2\pi)^3} \int \frac{d^3p}{-p_t} \sqrt{-g}\, f\, p^\mu p^\nu. \label{eq:Tmunu}
\end{align}
\end{subequations}
At equilibrium, these quantities can be related to the fluid $4$-velocity as follows:
\begin{subequations}\label{eq:macro_eq}
\begin{align}
 N^\mu_{\rm eq} =& n u^\mu, \label{eq:Nmu_eq}\\
 T^{\mu\nu}_{\rm eq} =& \epsilon u^\mu u^\nu +
 P \Delta^{\mu\nu}, \label{eq:Tmunu_eq}
\end{align}
\end{subequations}
where $n$, $\epsilon$ and $P$ are the particle number density, internal energy density and
isotropic pressure, while the projector $\Delta^{\mu\nu}$ is defined by:
\begin{equation}
 \Delta^{\mu\nu} = g^{\mu\nu} + u^\mu u^\nu.
\end{equation}

\section{Relativistic rotating gas}\label{sec:rot}

\subsection{Flow parameters} \label{sec:rot:flow}

Let us consider the problem of a relativistic gas in Minkowski space rotating about
the $z$ axis with constant angular velocity
$\bm{\omega} = (0,0,\omega)$. In special relativity, such a system can only exist up 
to a distance $\rho = \omega^{-1}$ from the rotation axis, where co-rotating particles 
rotate at the speed of light (SOL).

It is convenient to work in cylindrical coordinates
$(x^0, x^1, x^2, x^3) = (t, \rho, \varphi, z)$, with respect to which the
Minkowski line element takes the following form:
\begin{equation}
 ds^2 = -dt^2 + d\rho^2 + \rho^2d\varphi^2 + dz^2,\label{eq:ds2}
\end{equation}
from where the metric components can be read:
\begin{equation}
 g_{\mu\nu} = {\rm diag} (-1, 1, \rho^2, 1).\label{eq:gmunu}
\end{equation}
The $4$-velocity for such a flow can be written as \cite{cercignani02}:
\begin{equation}
 \beta u^\mu = \Omega^0 (1, 0, \omega, 0)^T, \label{eq:macro_cyl}
\end{equation}
where $\Omega^0$ is a constant. Keeping in mind that $u_\mu = g_{\mu\nu} u^\nu$,
it can be seen that the Killing equation \eqref{eq:killing} is automatically satisfied.

The requirement that the $4$-velocity has unit norm implies the following relation:
\begin{equation}
 -u_\mu u^\mu = \left(\frac{\Omega^0}{\beta}\right) (1 - \rho^2 \omega^2) = 1,
\end{equation}
such that the inverse temperature is given by:
\begin{equation}
 \beta \equiv \beta(\rho) = \beta_0 \sqrt{1 - \rho^2 \omega^2},
 \label{eq:beta_rotation}
\end{equation}
where $\beta_0$ is the inverse temperature on the rotation axis (i.e $\rho = 0$), and is linked to $\Omega^0$ as follows:
\begin{equation}
 \beta_0 = \Omega^0.
\end{equation}
Identifying $\rho \omega$ as the velocity of a particle rotating with angular velocity
$\omega$ at a distance $\rho$ from the rotation axis, the factor
\begin{equation}
 \gamma = \frac{1}{\sqrt{1 - \rho^2\omega^2}}\label{eq:gamma_def}
\end{equation}
can be seen to represent the corresponding Lorentz factor, such that $u^\mu$ can be written as:
\begin{equation}
 u^\mu = (\gamma, 0, \gamma \omega, 0)^T.\label{eq:umu}
\end{equation}
It can be seen that $\gamma \rightarrow \infty$ as the SOL is approached (i.e.~$\rho\omega \rightarrow 1$),
rendering the four-velocity \eqref{eq:umu} ill-defined past this point.
\subsection{Transformation matrix}\label{sec:rot:transf}
Let the matrices $L^\hmu{}_\mu$, $L^\mu{}_\hmu$ represent the transformation from the coordinate system
$\{x^\mu\}$ to a new coordinate system $\{x^\hmu\}$, such that
\begin{equation}
 L^{\hmu}{}_{\mu} = \frac{\partial x^{\hmu}}{\partial x^\mu}, \qquad
 L^{\mu}{}_{\hmu} = \frac{\partial x^{\mu}}{\partial x^\hmu},
\end{equation}
According to the algorithm described in Ref.~\cite{cardall15}, the above transformation
must be chosen such that the resulting flow $4$-velocity is
\begin{equation}
 u^{\hmu} = (1,0,0,0)^T\label{eq:uhat}
\end{equation}
and the resulting metric has the Minkowski form
\begin{equation}
 g_{\hmu\hnu} = g_{\mu\nu} L^\mu{}_\hmu L^\nu{}_\hnu = {\rm diag}(-1, 1,1, 1).
 \label{eq:gmunu_hat}
\end{equation}
The resulting coordinate system is in general non-holonomic and its associated tetrad
vectors and corresponding dual 1-forms are:
\begin{equation}
 e_{\hmu} = L^\mu{}_\hmu \partial_\mu, \qquad
 \omega^\hmu = L^\hmu{}_\mu dx^\mu, \label{eq:tetrad_L}
\end{equation}
such that
\begin{equation}
 \braket{\omega^\hmu, e_\hnu} \equiv \omega^\hmu_\lambda e^\lambda_\hnu = \delta^\hmu{}_\hnu.
 \label{eq:tetrad_ortho}
\end{equation}
From the above identifications, it can be seen that
\begin{equation}
 L^{\hmu}{}_{\mu}  L^{\mu}{}_{\hnu}  = \delta^{\hmu}{}_{\hnu}, \qquad
 L^{\mu}{}_{\hmu}  L^{\hmu}{}_{\nu}  = \delta^{\mu}{}_{\nu}.
\end{equation}
In this section, we reproduce the steps indicated in Ref.~\cite{cardall15}, specialising
to the problem of a rigidly rotating gas on Minkowski space.

Since in the new frame, the flow $4$-velocity has components given by Eq.~\eqref{eq:uhat},
the components of $L^\mu{}_\hmu$ are easily identified for $\hmu = \hat{t}$:
\begin{equation}
 L^\mu{}_{\hat{t}} = u^\mu,
\end{equation}
due to the transformation property:
\begin{equation}
 u^\mu = L^\mu{}_\hmu u^\hmu.
\end{equation}
Similarly, it is convenient to introduce three spatial vectors $w$, $y$ and $z$ of unit norm,
which have the following components in the comoving frame:
\begin{equation}
 w^\hmu = (0,1,0,0)^T, \qquad
 y^\hmu = (0,0,1,0)^T, \qquad
 z^\hmu = (0,0,0,1)^T.
\end{equation}
Thus, $L^\mu{}_{\hmu}$ can be written as:
\begin{equation}
 L^\mu{}_{\hmu} = (u^\mu, w^\mu, y^\mu, z^\mu) \equiv
 \begin{pmatrix}
  u^0 & w^0 & y^0 & z^0\\
  u^1 & w^1 & y^1 & z^1\\
  u^2 & w^2 & y^2 & z^2\\
  u^3 & w^3 & y^3 & z^3
 \end{pmatrix},
\end{equation}
while its inverse $L^\hmu{}_\mu$ is given by:
\begin{equation}
 L^\hmu{}_{\mu} = (-u_\mu, w_\mu, y_\mu, z_\mu)^T \equiv
 \begin{pmatrix}
  -u_0 & -u_1 & -u_2 & -u_3\\
  w_0 & w^1 & w_2 & w_3\\
  y_0 & y_1 & y^2 & y_3\\
  z_0 & z_1 & z_2 & z_3
 \end{pmatrix}.
\end{equation}

Let us now define the normal $n_\mu = (-1, 0, 0, 0)$ to the three-surface of equal $t$.
It is convenient to split $u^\mu$ as follows:
\begin{equation}
 u^\mu = \Lambda(n^\mu + v^\mu),
\end{equation}
where $v$ is orthogonal to $n$, i.e.~$v^\mu n_\mu = 0$. Taking the contraction of the above expression
with $n_\mu$ gives:
\begin{equation}
 \Lambda = \gamma,
\end{equation}
hence
\begin{equation}
 v^\mu = (0, 0, \omega, 0),
\end{equation}
in accord with Eq.~(69) of Ref.~\cite{cardall15}.
Next, the vectors $w^\mu$, $y^\mu$ and $z^\mu$ can be written as:
\begin{equation}
 w^\mu = An^\mu + a^\mu, \qquad
 y^\mu = Bn^\mu + b^\mu, \qquad
 z^\mu = Cn^\mu + c^\mu,
\end{equation}
where $a$, $b$ and $c$ are orthogonal to $n$:
\begin{equation}
 n_\mu a^\mu = n_\mu b^\mu = n_\mu c^\mu = 0.\label{eq:abc_n}
\end{equation}
Since in our case, $n_\mu = (-1, 0, 0, 0)$, the above equations imply:
\begin{equation}
 a^0 = b^0 = c^0 = 0.\label{eq:abc_0}
\end{equation}

Using Eq.~\eqref{eq:gmunu_hat} for $\hmu = \hatt$ and $\hnu = \hati$ shows that
$w$, $y$ and $z$ are orthogonal to $u$:
\begin{equation}
 u_\mu w^\mu = u_\mu y^\mu = u_\mu z^\mu = 0,
\end{equation}
from where the following relations can be found:
\begin{equation}
 v_i a^i = A, \qquad v_i b^i = B, \qquad v_i c^i = C.\label{eq:abc_u}
\end{equation}
Specialising to the metric \eqref{eq:gmunu} and using the four-flow \eqref{eq:umu}
gives:
\begin{subequations}\label{eq:ABC}
\begin{align}
 A =& \rho^2\omega a^2, \label{eq:A}\\
 B =& \rho^2\omega b^2, \label{eq:B}\\
 C =& \rho^2\omega c^2. \label{eq:C}
\end{align}
\end{subequations}

Finally, using Eqs.~\eqref{eq:gmunu_hat} for $\hmu = \hati$ and $\hnu = \hatj$
for $\hati \neq \hatj$ shows that $w$, $y$ and $z$ are mutually orthogonal:
\begin{subequations}\label{eq:wyz_ortho}
\begin{align}
 w_\mu y^\mu = a_i b^i - AB = 0, \label{eq:wy_ortho}\\
 w_\mu z^\mu = a_i c^i - AC = 0, \label{eq:wz_ortho}\\
 y_\mu z^\mu = b_i c^i - BC = 0,\label{eq:yz_ortho}
\end{align}
\end{subequations}
while for $\hati = \hatj$, the following unit norm conditions are recovered:
\begin{subequations}\label{eq:wyz_norm}
\begin{align}
 w_\mu w^\mu = a_ia^i - A^2 = 1, \label{eq:w_norm}\\
 y_\mu y^\mu = b_ib^i - B^2 = 1, \label{eq:y_norm}\\
 z_\mu z^\mu = c_ic^i - C^2 = 1.\label{eq:z_norm}
\end{align}
\end{subequations}
Eqs.~\eqref{eq:abc_0}, \eqref{eq:abc_u}, \eqref{eq:wyz_ortho} and \eqref{eq:wyz_norm}
provide $12$ constraints for the $12$ components of $a$, $b$ and $c$ and the
$3$ parameters $A$,$B$ and $C$. This leaves us with three unspecified constraints, which
amount to the freedom of an $SO(3)$ rotation of the tetrad field\footnote{The other $3$ degrees of
freedom associated with the $SO(3,1)$ Lorentz invariance of the tetrad field are fixed
by imposing Eq.~\eqref{eq:uhat}.}.
Eqs.~(B1) and (B2) of Ref.~\cite{cardall15} suggest
fixing $b^1$, $c^1$ and $c^2$ to $0$:
\begin{equation}
 b^1 = c^1 = c^2 = 0.
\end{equation}
Eq.~\eqref{eq:C} can now be used to find that $C = 0$, while Eq.~\eqref{eq:z_norm} allows us to
fix $c^3 = 1$. Furthermore, Eqs.~\eqref{eq:wz_ortho} and \eqref{eq:yz_ortho} imply that
$a^3 = b^3 = 0$. Moreover, Eqs.~\eqref{eq:B} and \eqref{eq:y_norm} can be used to find that
$b^2 = \rho^{-1} \gamma$ and $B = \rho\omega\gamma$. Combining Eqs.~\eqref{eq:wy_ortho} and
\eqref{eq:A} shows that $a^2 = 0$, while Eq.~\eqref{eq:w_norm} can be used to fix $a^1 = 1$.
Thus, the following expressions are obtained for $w$, $y$ and $z$:
\begin{subequations}
\begin{align}
 w^\mu =& (0, 1, 0, 0)^T, & w_\mu =& (0, 1, 0, 0), \\
 y^\mu =& (\beta\gamma, 0, \rho^{-1}\gamma, 0)^T, & y_\mu =& (-\beta\gamma, 0, \rho\gamma, 0), \\
 z^\mu =& (0, 0, 0, 1)^T, & z_\mu =& (0, 0, 0, 1),
\end{align}
\end{subequations}
giving rise to the following transformation matrices:
\begin{equation}
 L^\mu{}_{\hmu} =
 \begin{pmatrix}
  \gamma & 0 & \rho\omega\gamma & 0\\
  0 & 1 & 0 & 0\\
  \omega \gamma & 0 & \rho^{-1} \gamma & 0\\
  0 & 0 & 0 & 1
 \end{pmatrix}, \qquad
 L^\hmu{}_{\mu} =
 \begin{pmatrix}
  \gamma & 0 & -\rho^2\omega\gamma & 0\\
  0 & 1 & 0 & 0\\
  -\rho\omega\gamma & 0 & \rho \gamma & 0\\
  0 & 0 & 0 & 1
 \end{pmatrix}.\label{eq:L_final}
\end{equation}
%
\subsection{Tetrad}\label{sec:rot:tetrad}
The tetrad frame vectors $e_\hmu$ and co-frame 1-forms $\omega^\hmu$ can be obtained by substituting
Eqs.~\eqref{eq:L_final} into Eqs.~\eqref{eq:tetrad_L}:
\begin{subequations} \label{eq:tetrad}
\begin{align}
 e_\hatt =& \gamma\partial_t + \omega\gamma \partial_\varphi,&
 \omega^\hatt =& \gamma dt - \rho^2 \omega \gamma d\varphi,\\
 e_{\hat{\rho}} =& \partial_\rho,&
 \omega^{\hat{\rho}} =& d\rho,\\
 e_{\hat{\varphi}} =& \rho\omega\gamma\partial_t + \rho^{-1}\gamma \partial_\varphi,&
 \omega^{\hat{\varphi}} =& -\rho\omega\gamma dt+ \rho\gamma d\varphi,\\
 e_{\hat{z}} =& \partial_z,&
 \omega^{\hat{z}} =& dz.
\end{align}
\end{subequations}
It can be checked that the orthogonality relation \eqref{eq:tetrad_ortho} is automatically satisfied.

Since the metric expressed with respect to this non-holonomic tetrad \eqref{eq:tetrad} is the Minkowski metric,
the connection coefficients are given by \cite{misner73}:
\begin{equation}
 \Gamma_{\hmu\hnu\hlambda} = \frac{1}{2} (c_{\hmu\hnu\hlambda} +
 c_{\hmu\hlambda\hnu} - c_{\hnu\hlambda\hmu}),
 \label{eq:conn_def}
\end{equation}
where the Cartan coefficients are defined as:
\begin{equation}
 c_{\hmu\hnu}{}^{\hlambda} = \braket{\omega^{\hlambda}, [e_{\hmu}, e_{\hnu}]}.
 \label{eq:cartan_def}
\end{equation}
Due to the anti-symmetry of the commutator, the Cartan coefficients are antisymmetric with respect
to the first two indices:
\begin{equation}
 c_{\hmu\hnu}{}^{\hlambda} = -c_{\hnu\hmu}{}^{\hlambda}.\label{eq:cartan_asym}
\end{equation}
We note that this also implies an anti-symmetry in the first two indices of the connection coefficients:
\begin{equation}
 \Gamma_{\hmu\hnu\hlambda} = -\Gamma_{\hnu\hmu\hlambda}.\label{eq:conn_asym}
\end{equation}

To calculate the commutators of the basis vectors $e_{\hmu}$ in Eq.~\eqref{eq:cartan_def},
the following property can be used:
\begin{equation}
 \partial_\rho \gamma = \rho\omega^2 \gamma^3.
\end{equation}
The non-vanishing Cartan coefficients are:
\begin{equation}
 c_{\hatt\hat{\rho}\hatt} = \rho\omega^2\gamma^2, \qquad
 c_{\hat{\rho}\hat{\varphi}\hatt} = -2\omega\gamma^2, \qquad
 c_{\hat{\rho}\hat{\varphi}\hat{\varphi}} = -\rho^{-1} \gamma^2.\label{eq:cartan}
\end{equation}
Hence, the connection coefficients are:
\begin{gather}
 \Gamma_{\hatt\hat{\rho}\hatt} = \rho\omega^2\gamma^2, \quad
 \Gamma_{\hatt\hat{\rho}\hat{\varphi}} = \omega\gamma^2, \quad
 \Gamma_{\hatt\hat{\varphi}\hat{\rho}} = -\omega\gamma^2, \quad
 \Gamma_{\hat{\rho}\hat{\varphi}\hatt} = -\omega\gamma^2, \quad
 \Gamma_{\hat{\rho}\hat{\varphi}\hat{\varphi}} = -\rho^{-1}\gamma^2.
 \label{eq:conn}
\end{gather}

\subsection{Momentum space}\label{sec:rot:mspace}

Following the application of the transformation $L^\mu{}_\hmu$, the momentum space also changes
according to
\begin{equation}
 p^\hmu = L^\hmu{}_\mu p^\mu,
\end{equation}
i.e.~such that $p^\hmu e_{\hmu} = p^\mu \partial_\mu$. For completeness, we list the
components of $p^\hmu$ in terms of the cylindrical components $p^\mu$:
\begin{equation}
 p^\hmu = (\gamma p^t - \rho^2\omega \gamma p^\varphi, p^\rho, -\rho\omega\gamma p^t +
 \rho\gamma p^\varphi, p^z)^T.
\end{equation}
It can be checked that the above components obey the mass-shell condition with respect to
the Minkowski metric:
\begin{equation}
 \eta_{\hmu\hnu} p^\hmu p^\hnu = g_{\mu\nu} p^\mu p^\nu = -m^2.
\end{equation}

As pointed out in Ref.~\cite{romatschke11} for lattice Boltzmann simulations, a separation of
variables with respect to the spherical coordinate system is convenient to construct quadratures
for the evaluation of the moments in Eqs.~\eqref{eq:hydro}.
Following Ref.~\cite{cardall15}, we can introduce a metric for the spatial part of the momentum space, such that
the line element is given by:
\begin{equation}
 d\Phi^2 = \delta_{\hati\hatj} dp^{\hati} dp^{\hatj}.
\end{equation}
Changing to the spherical coordinates in momentum space $p^{\ti} = (p, \theta_p, \varphi_p)$,
defined through:
\begin{subequations}\label{eq:tetrad_mom_sph}
\begin{align}
 p^{\hat{\rho}} =& p \sin\theta_p \cos\varphi_p, \\
 p^{\hat{\varphi}} =& p \sin\theta_p \sin\varphi_p, \\
 p^{\hat{z}} =& p \cos\theta_p,
\end{align}
\end{subequations}
induces a new metric $\lambda_{\ti\tj}$, defined as:
\begin{equation}
 \lambda_{\ti\tj} = \delta_{\hati\hatj} P^\hati{}_\ti P^\hatj{}_\tj,
 \label{eq:lambda_def}
\end{equation}
where $P^\hati{}_\ti$ and its inverse, $P^\ti{}_\hati$, represent
the matrices for the transformation between the coordinates $p^\ti$ and $p^\hati$,
being defined as:
\begin{equation}
 P^\hati{}_\ti = \frac{\partial p^\hati}{\partial p^\ti}, \qquad
 P^\ti{}_\hati = \frac{\partial p^\ti}{\partial p^\hati}, \qquad
\end{equation}
The components of $P^\hati{}_\ti$ can be calculated from Eqs.~\eqref{eq:tetrad_mom_sph}:
\begin{equation}
 P^\hati{}_\ti =
 \begin{pmatrix}
  \sin\theta_p \cos\varphi_p & p\cos\theta_p \cos\varphi_p & -p\sin\theta_p \sin\varphi_p\\
  \sin\theta_p \sin\varphi_p & p\cos\theta_p \sin\varphi_p & p\sin\theta_p \cos\varphi_p\\
  \cos\theta_p & -p\sin\theta_p & 0
 \end{pmatrix},
\end{equation}
while its inverse is given by:
\begin{equation}
 P^\ti{}_\hati =
 \begin{pmatrix}
  \sin\theta_p \cos\varphi_p & \sin\theta_p \sin\varphi_p & \cos\theta_p \\
   \frac{1}{p}\cos\theta_p \cos\varphi_p & \frac{1}{p} \cos\theta_p \sin\varphi_p & -\frac{1}{p} \sin\theta_p\\
  -\frac{\sin\varphi_p}{p\sin\theta_p} & \frac{\cos\varphi_p}{p\sin\theta_p} & 0
 \end{pmatrix}.
\end{equation}
Hence, the new metric $\lambda_{\ti\tj}$ takes the following form:
\begin{equation}
 \lambda_{\ti\tj} = {\rm diag}(1, p^2, p^2\sin^2\theta_p).\label{eq:lambda}
\end{equation}

Integration in momentum space can now be performed using the following measure:
\begin{equation}
 \frac{d^3\widetilde{p}}{-p_\hatt} \frac{\sqrt{\lambda} }{(2\pi)^3},
 \label{eq:dptilda}
\end{equation}
where $d^3\widetilde{p} = dp\,d\theta_p\,d\varphi_p$ and
$\sqrt{\lambda} = p^2 \sin\theta_p$ is the square root of the determinant of $\lambda_{\ti\tj}$ \eqref{eq:lambda}.
The hydrodynamic moments \eqref{eq:hydro} can now be computed:
\begin{subequations}\label{eq:hydro_tetrad}
\begin{align}
 N^\hmu =& \frac{1}{(2\pi)^3} \int_{0}^\infty \frac{dp\, p^2}{\sqrt{p^2 + m^2}} \int d \Omega_p \, f\, p^\hmu, \label{eq:Nmu_tetrad}\\
 T^{\mu\nu} =& \frac{1}{(2\pi)^3} \int_{0}^\infty \frac{dp\, p^2}{\sqrt{p^2 + m^2}} \int d \Omega_p \, f\, p^\mu p^\nu, \label{eq:Tmunu_tetrad}
\end{align}
\end{subequations}
where $d\Omega_p = \sin\theta_p d\theta_p d\varphi_p$ is the elementary solid angle in momentum space.

The tetrad formalism presented in this section can be employed for 
the computation of the moments of $f$ with respect to spherical coordinates
in momentum space in a manner which is decoupled from the background space-time.
In lattice Boltzmann simulations, this decoupling can facilitate the use of 
the quadrature methods developed for Minkowski space when arbitrary space-times are considered.

For completeness, we also consider the momentum decomposition \cite{cardall15} of the transformation
matrix $P^\ti{}_\hati$, but since this topic is not relevant to the remainder of this paper, we
present it in appendix~\ref{app:momdec}.

\subsection{Boltzmann equation and equilibrium states}\label{sec:rot:boltz}

Owing to the general covariance of the Boltzmann equation \eqref{eq:boltz}, the transition to
the new coordinates in momentum space is straightforward. However, it is convenient to
express the Boltzmann equation with respect to the original space-time coordinates, since
the relation between the coordinates $x^\hmu$ in the new system and the coordinates $x^\mu$
in the original system can be found by integrating the system of equations given by
$\partial x^\hmu / \partial x^\mu = L^\hmu{}_\mu$. Thus, we write the Boltzmann equation as:
\begin{equation}
 p^\hmu L^\mu{}_\hmu \frac{\partial f}{\partial x^\mu} - \Gamma^{\hati}{}_{\hmu\hnu} p^\hmu p^\hnu
 P^\ti{}_\hati \frac{\partial f}{\partial p^\ti} = C[f],\label{eq:boltz_tetrad}
\end{equation}
where $f \equiv f(x^\mu, p^\ti)$ depends on the original coordinates $x^\mu$ and the
new momentum space variables $p^\ti = (p, \theta_p, \varphi_p)$.

To study the equilibrium hydrodynamic profiles with respect to the new frame, the equilibrium distribution
function \eqref{eq:feq} can be written as:
\begin{equation}
 \feq = Z \left[\exp\left(\beta p^\hatt\right) - \varepsilon\right]^{-1},\label{eq:feq_tetrad}
\end{equation}
where the macroscopic velocity with respect to the tetrad has components $u^\hmu = (1, 0, 0, 0)$.
The restrictions on $\beta$ can be found by substituting Eq.~\eqref{eq:feq_tetrad} into
the Boltzmann equation:
\begin{equation}
 p^\hatt p^\hmu L^\mu{}_\hmu \partial_\mu \beta =
 -\beta \rho\omega^2 \gamma^2 p^{\hat{\rho}} p^\hatt,
\end{equation}
where Eqs.~\eqref{eq:conn} were used for the connection coefficients. The solution of the above equation is:
\begin{equation}
 \beta = \frac{\beta_0}{\gamma},
\end{equation}
where $\gamma$ is defined in Eq.~\eqref{eq:gamma_def} and $\beta_0$ represents the inverse temperature
at $\rho = 0$. The above result is in agreement with Eq.~\eqref{eq:beta_rotation}.

\subsection{Equilibrium hydrodynamic profiles}\label{sec:rot:eq}

The hydrodynamic fields can be found by integrating Eq.~\eqref{eq:feq_tetrad} over the momentum space, using
the integration measure in Eq.~\eqref{eq:dptilda}:
\begin{subequations}
\begin{align}
 N^\hmu =& \frac{1}{(2\pi)^3} \int_0^\infty \frac{dp}{p^\hatt}\, p^2\, \feq \int d\Omega_p\, p^\hmu,\\
 T^{\hmu\hnu} =& \frac{1}{(2\pi)^3} \int_0^\infty \frac{dp}{p^\hatt}\, p^2\, \feq \int d\Omega_p\, p^\hmu p^\hnu.
\end{align}
\end{subequations}
Due to the symmetries of the integration measure, it can be seen that $N^{\hati} = 0$, $T^{\hatt\hati} = 0$ and
$T^{\hati\hatj} \sim \delta^{\hati\hatj}$, such that
\begin{equation}
 N^\hmu = (n, 0, 0, 0), \qquad
 T^{\hmu\hnu} = {\rm diag}(\epsilon, P, P, P),
\end{equation}
where
\begin{subequations}\label{eq:macro_sph}
\begin{align}
 n =& \frac{1}{2\pi^2} \int_0^\infty dp\, p^2\, \feq,\\
 \epsilon =& \frac{1}{2\pi^2} \int_0^\infty dp\, p^\hatt\, p^2\, \feq,\\
 P =& \frac{1}{6\pi^2} \int_0^\infty \frac{dp}{p^\hatt}\, p^3\, \feq.
\end{align}
\end{subequations}

Expressions for $n$, $\epsilon$ and $P$ can be derived for all three distributions, the expression
\eqref{eq:feq} can be written in a power series, as follows:
\begin{equation}
 \feq (Z; \beta, u)
 = \sum_{j = 0}^\infty \varepsilon^j  \feq_{\rm M-J}[Z; \beta(j + 1), u],
 \label{eq:feq_MJ}
\end{equation}
where $\feq_{\rm M-J}$ is the M-J distribution given by Eq.~\eqref{eq:feq} for
$\varepsilon = 0$. Hence, the hydrodynamic profiles for the F-D and
B-E distributions can be determined by summing over the M-J
profiles at increasing values of $\beta$. 
Switching to the variable $x = p^\hatt / m = \sqrt{1 + p^2/m^2}$ 
in Eqs.~\eqref{eq:macro_sph} allows $n$, $\epsilon$ and $p$
to be written as:
\begin{subequations}
\begin{align}
 n_{\rm M-J} =& \frac{Z m^3}{2\pi^2} \int_1^\infty dx\, x\sqrt{x^2-1} e^{-\beta m x},\\
 \epsilon_{\rm M-J} =& \frac{Z m^4}{2\pi^2} \int_1^\infty dx\, x^2\sqrt{x^2-1} e^{-\beta m x},\\
 P_{\rm M-J} =& \frac{Z m^3 }{6\pi^2} \int_1^\infty dx\, (x^2-1)^{3/2} e^{-\beta m x}.
\end{align}
\end{subequations}
The integrals above can be calculated exactly using the integral expression for the modified Bessel
functions of the second kind \cite{nist}:
\begin{equation}
 K_\nu(z) = \frac{\sqrt{\pi} (z / 2)^\nu}{\Gamma(\nu + 1/2)} \int_1^\infty dt\, e^{-zt}
 (t^2 - 1)^{\nu - 1/2},
\end{equation}
together with the following recurrence relation \cite{nist}:
\begin{equation}
 K_{\nu \pm 1}(z) = -K_\nu'(z) \pm \frac{\nu}{z} K_\nu(z).
\end{equation}
The following expressions are obtained for $n$, $\epsilon$ and $P$:
\begin{subequations}\label{eq:hydro_sums}
\begin{align}
 n(\beta) =& \sum_{j = 0}^\infty \varepsilon^j n_{\rm M-J}[(j + 1)\beta], \\
 \epsilon(\beta) =& \sum_{j = 0}^\infty \varepsilon^j \epsilon_{\rm M-J}[(j + 1)\beta],\label{eq:epsilon_sumj} \\
 P(\beta) =& \sum_{j = 0}^\infty \varepsilon^j p_{\rm M-J}[(j + 1)\beta],\label{eq:P_sumj}
\end{align}
\end{subequations}
written in terms of the corresponding expressions calculated using the M-J distribution
at inverse temperature $(j + 1)\beta$:
\begin{subequations}\label{eq:hydro_MJ}
\begin{align}
 n_{\rm M-J} =& \frac{Z}{\pi^2 \beta^3} \widetilde{K}_2(m \beta),\\
 \epsilon_{\rm M-J} =& \frac{3Z}{\pi^2 \beta^4} \left[\widetilde{K}_2(m \beta)
 + \frac{(m\beta)^2}{6} \widetilde{K}_1(m \beta)\right],\label{eq:hydro_MJ_e}\\
 P_{\rm M-J} =& \frac{Z}{\pi^2 \beta^4} \widetilde{K}_2(m \beta),\label{eq:hydro_MJ_P}
\end{align}
\end{subequations}
where
\begin{equation}
 \widetilde{K}_\nu(m\beta) \equiv \frac{(\beta m)^\nu}{\Gamma(\nu + 1)} K_\nu(\beta m)
\end{equation}
reduces to unity in the massless limit (i.e.~$m\rightarrow 0$).
Substituting Eqs.~\eqref{eq:hydro_MJ_e} and \eqref{eq:hydro_MJ_P} into Eqs.~\eqref{eq:epsilon_sumj} and 
\eqref{eq:P_sumj}, respectively, gives the following expression for the equation of state $w = P / \epsilon$:
\begin{align}
 w =& \frac{\sum_{j = 0}^\infty 
 \frac{\varepsilon^j}{(j+1)^4} \widetilde{K}_2[\gamma^{-1} m\beta_0(j+1)]}
 {\sum_{j = 0}^\infty \frac{\varepsilon^j}{(j+1)^4} \left\{\widetilde{K}_2[\gamma^{-1} m\beta_0(j+1)]
 + \frac{1}{6}\left[\gamma^{-1} m\beta_0(j+1)\right]^2 \widetilde{K}_1[\gamma^{-1} m\beta_0(j+1)]\right\}}\nonumber\\
 =& \frac{1}{3}\left\{1 + \frac{1}{2}(\gamma^{-1} m\beta_0)^2
 \frac{\sum_{j = 0}^\infty \frac{\varepsilon^j}{(j+1)^2} \widetilde{K}_1[\gamma^{-1} m\beta_0(j+1)]}
 {\sum_{j = 0}^\infty \frac{\varepsilon^j}{(j+1)^4} \widetilde{K}_2[\gamma^{-1} m\beta_0(j+1)]}\right\}^{-1},
 \label{eq:w}
\end{align}
where $\beta_0 = \beta(\rho = 0)$ is the inverse temperature on the rotation axis.
It can be seen that $w$ only depends on two parameters: $\beta_0 m$ and $\rho\omega$.

\subsection{Analysis of the equilibrium hydrodynamic profiles}

Due to the dependence \eqref{eq:beta_rotation} of the inverse temperature $\beta$ depends on the position 
$\rho$, the hydrodynamic variables $n$, $\epsilon$ and $P$ diverge as the speed of light surface (SOL) is approached 
$\rho \rightarrow \omega^{-1}$, for all three statistics considered. To further investigate 
their properties, approximations or numerical techniques can be employed. 
For the M-J equilibrium distribution, Eqs.~\eqref{eq:hydro_MJ} give analytic closed-form
expressions for $n$, $\epsilon$ and $P$. For the F-D and B-E cases, closed-form expressions
are, to the best of our knowledge, not known for general values of the mass $m$. 
In subsubsection~\ref{sec:rot:an:smallm}, an analysis of the small $m\beta$ limit is presented,
while in subsubsection~\ref{sec:rot:an:num}, we present some numerical results.

\subsubsection{Small mass or large temperature}\label{sec:rot:an:smallm}

At small $m\beta$, the modified Bessel functions in Eqs.~\eqref{eq:hydro_MJ} can be expanded as \cite{nist}:
\begin{equation} \label{eq:expandK}
 \widetilde{K}_{2}(m\beta) = 1 - \frac{m\beta^2}{4} + O[(m\beta)^3], \qquad
 \widetilde{K}_{1}(m\beta) = 1 + O[(m\beta)],
\end{equation}
such that the first mass correction in Eqs.~\eqref{eq:hydro_MJ} can be obtained:
\begin{subequations}\label{eq:hydro_MJ_smallm}
\begin{align}
 n_{\rm M-J} =& \frac{Z}{\pi^2 \beta^3} \left\{1 - \frac{(m \beta)^2}{4} + O[(m\beta)^3]\right\},\\
 \epsilon_{\rm M-J} =& \frac{3Z}{\pi^2 \beta^4} \left\{1 - \frac{(m\beta)^2}{12} + O[(m\beta)^3]\right\}
 \label{eq:epsilon_MJ_smallm},\\
 P_{\rm M-J} =& \frac{Z}{\pi^2 \beta^4} \left\{1 - \frac{(m \beta)^2}{4} + O[(m\beta)^3]\right\}.
\end{align}
\end{subequations}
To obtain similar small mass corrections when the F-D and B-E statistics are employed,
the above results can be substituted into Eqs.~\eqref{eq:hydro_sums}.
It can be seen that continuing the above expansions to higher powers in $m\beta$ eventually will lead 
to divergent sums over $j$, since in Eqs.~\eqref{eq:hydro_sums}, $\beta$ is multiplied by $j+1$.
The reason for this apparent divergence is that the expansions in Eq.~\eqref{eq:expandK} are only 
valid for small arguments of the modified Bessel functions. However, it is still possible to obtain the 
corrections in $m\beta$ for the terms where the sums over $j$ do not diverge. 
In such terms, the sums over $j$ can be computed in terms of the Riemann zeta function 
$\zeta(z)$, which satisfies the following properties \cite{gradshteyn}:
\begin{equation}
 \sum_{j = 0}^\infty \frac{1}{(j+1)^z} = \zeta(z), 
 \sum_{j = 0}^\infty \frac{(-1)^j}{(j+1)^z} = (1 - 2^{1-z}) \zeta(z), 
\end{equation}
where $z > 1$. The values of $z$ which are of interest in the present case are \cite{gradshteyn}:
\begin{equation}
 \zeta(2) = \frac{\pi^2}{6}, \qquad \zeta(3) \simeq 1.20206, \qquad \zeta(4) = \frac{\pi^4}{90}.
\end{equation}
Thus, the small $m\beta$ limit of Eqs.~\eqref{eq:hydro_sums} when B-E statistics are employed 
reduces to:
\begin{subequations}\label{eq:hydro_BE_smallm}
\begin{align}
 n_{\rm B-E} =& \frac{\zeta(3)Z}{\pi^2 \beta^3} \left\{ 1+ O[(m\beta)^2]\right\},\\
 \epsilon_{\rm B-E} =& \frac{\pi^2Z}{30 \beta^4} \left\{1  - \frac{5}{4\pi^2}(m\beta)^2+O[(m\beta)^3]\right\},
 \label{eq:epsilon_BE_smallm}\\
 P_{\rm B-E} =&  \frac{\pi^2 Z}{90 \beta^4} \left\{ 1 - \frac{15}{4\pi^2}(m\beta)^2 + O[(m\beta)^3]\right\},
\end{align}
\end{subequations}
No correction can be obtained for $n_{\rm B-E}$ due to the divergent behaviour of $\zeta(1)$ \cite{gradshteyn}.
Similar expressions can be obtained for the F-D statistics:
\begin{subequations}\label{eq:hydro_FD_smallm}
\begin{align}
 n_{\rm F-D} =& \frac{3\zeta(3) Z}{4\pi^2 \beta^3} \left\{1- \frac{\ln(2)}{3\zeta(3)}(m\beta)^2 + O[(m\beta)^3]\right\},\\
 \epsilon_{\rm F-D} =& \frac{7\pi^2 Z}{240 \beta^4} \left\{1  - \frac{5}{\pi^2}(m\beta)^2+O[(m\beta)^3]\right\},
 \label{eq:epsilon_FD_smallm}\\
 P_{\rm F-D} =&  \frac{7\pi^2 Z}{720 \beta^4} \left\{ 1 - \frac{15}{\pi^2}(m\beta)^2+O[(m\beta)^3]\right\},
\end{align}
\end{subequations}
where the correction in $n_{\rm F-D}$ was obtained using the following identity \cite{gradshteyn}:
\begin{equation}
 \sum_{j = 0}^\infty \frac{(-1)^j}{j + 1} = \ln 2.
\end{equation}
Having obtained approximate expressions for the energy density and pressure, the equation of state $w$ \eqref{eq:w}
can be computed at the same order:
\begin{subequations}\label{eq:w_smallmb}
\begin{align}
 w_{\rm M-J} \equiv \frac{P_{\rm M-J}}{\epsilon_{\rm M-J}} =& \frac{1}{3} \left\{ 1 - \frac{1}{6}(\beta m)^2 + O[(m \beta m)^3] \right\},
 \label{eq:w_smallmb_MJ}\\
 w_{\rm B-E} \equiv \frac{P_{\rm B-E}}{\epsilon_{\rm B-E}} =& \frac{1}{3} \left\{ 1 - \frac{5}{2\pi^2} (\beta m)^2 + O[(m \beta)^3] \right\},
 \label{eq:w_smallmb_BE}\\
 w_{\rm F-D} \equiv \frac{P_{\rm F-D}}{\epsilon_{\rm F-D}} =& \frac{1}{3} \left\{ 1 - \frac{10}{\pi^2} (\beta m)^2 + O[(m \beta)^3] \right\}.
 \label{eq:w_smallmb_FD}
\end{align}
\end{subequations}
Since the equations of state only depend on the combination $m\beta$, the small mass and
high temperature limits coincide. For vanishing mass, Eqs.~\eqref{eq:w_smallmb} reduce to $w = 1/3$. 
This is also the case when the SOL is approached (i.e.~$\rho\omega \rightarrow 1$) and 
$\beta \rightarrow 0$, as can be seen from Figure~\ref{fig:plots} (c). We also note that,
as the SOL is approached, $n_{\rm M-J}$, $\epsilon_{\rm M-J}$ and $P_{\rm M-J}$ diverge as powers 
of $\gamma = (1 - \rho^2\omega^2)^{-1/2}$, as illustrated for $\epsilon_{\rm M-J}$ 
in Figure~\ref{fig:plots}(d).

\subsubsection{Numerical results}\label{sec:rot:an:num}

\begin{figure}[t]
\begin{center}
\begin{tabular}{cc}
\includegraphics[width=0.45\linewidth]{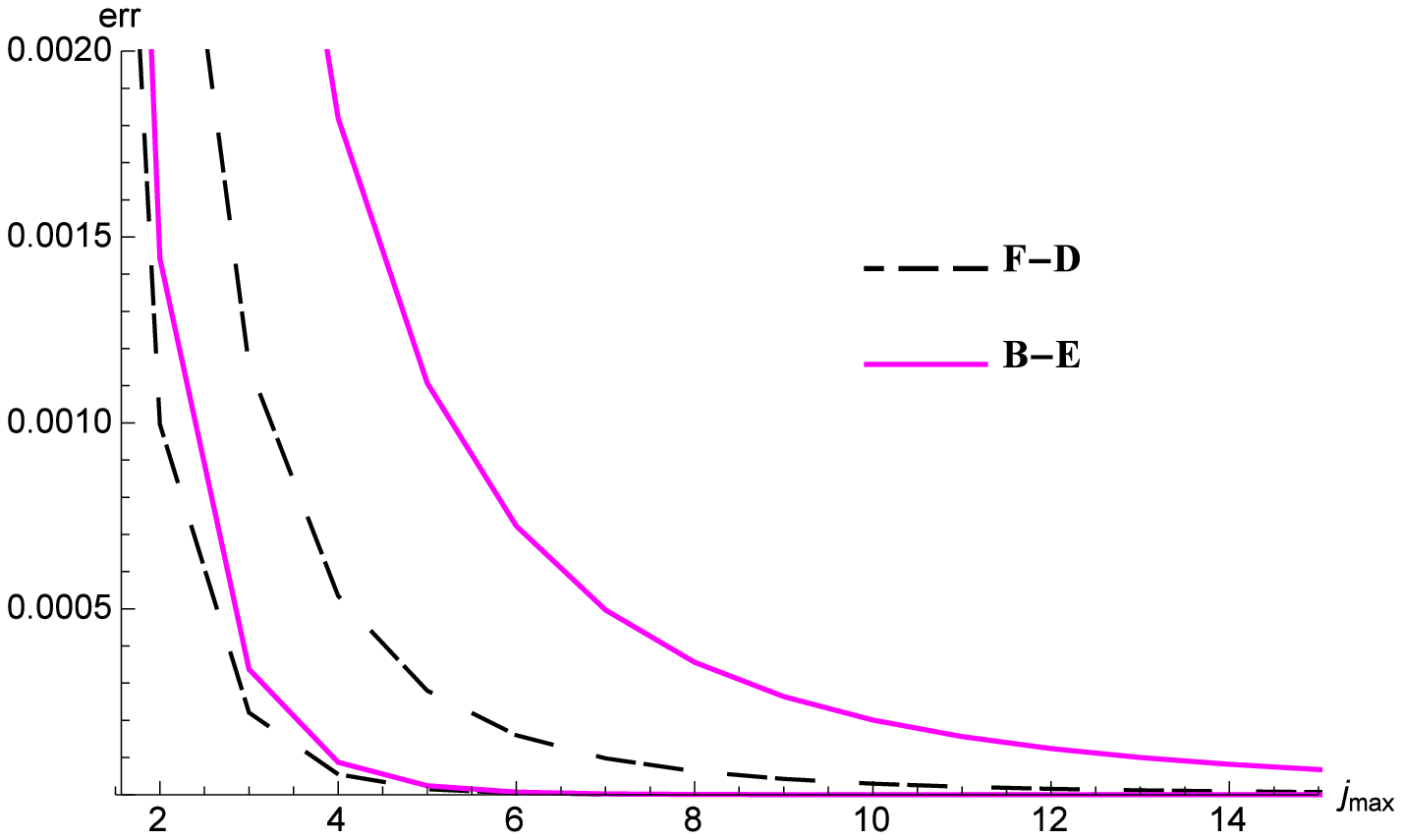} &
\includegraphics[width=0.45\linewidth]{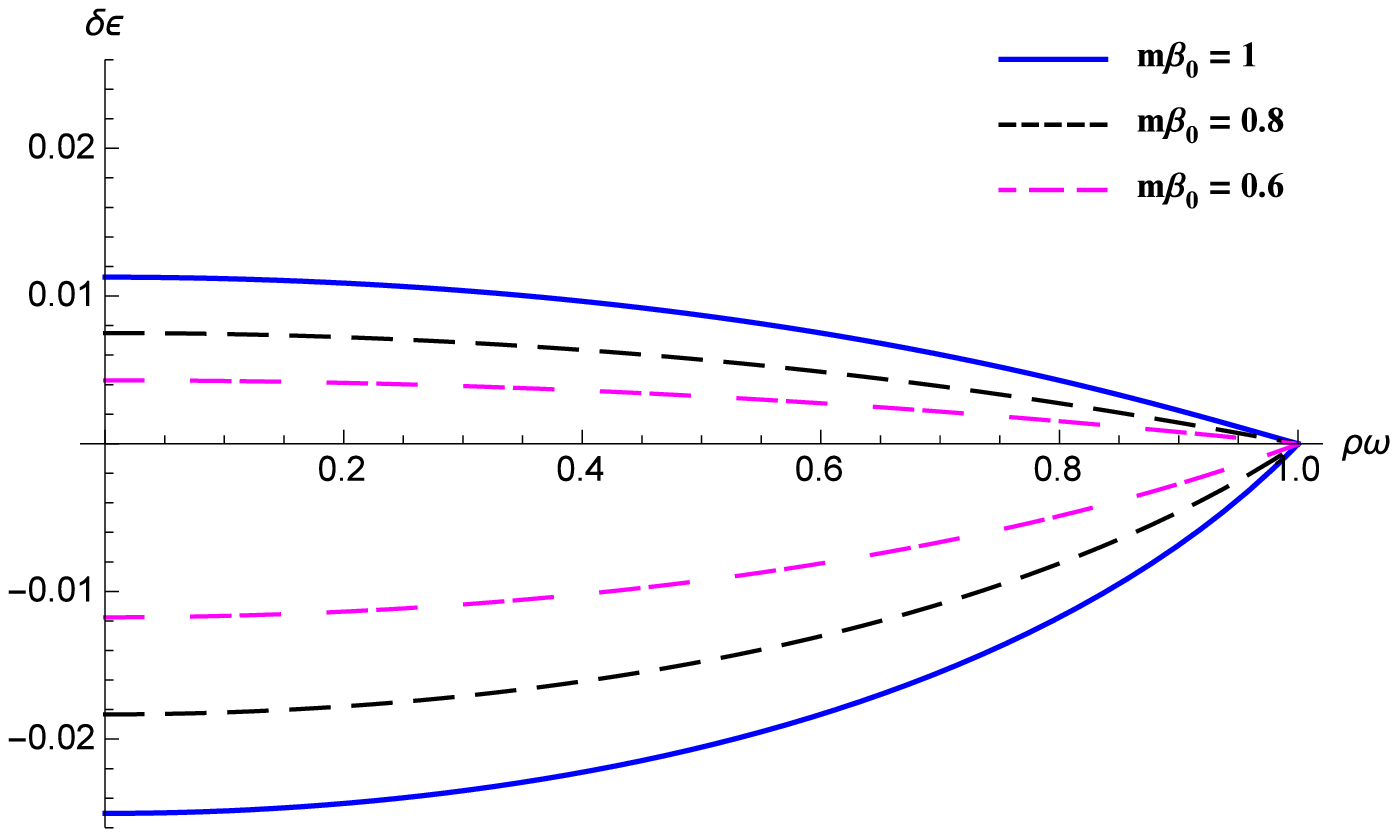} \\
(a) & (b)\\
\includegraphics[width=0.45\linewidth]{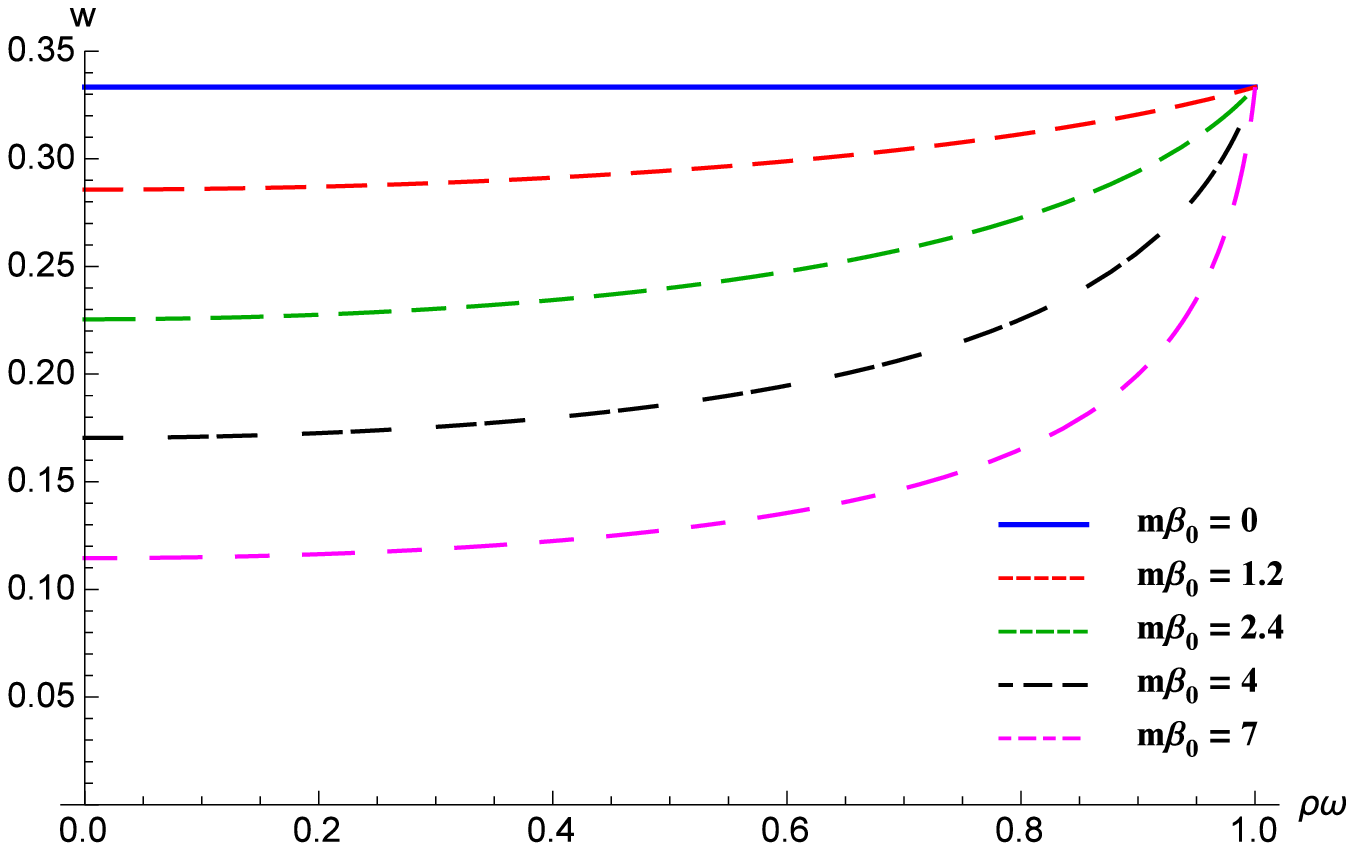} &
\includegraphics[width=0.45\linewidth]{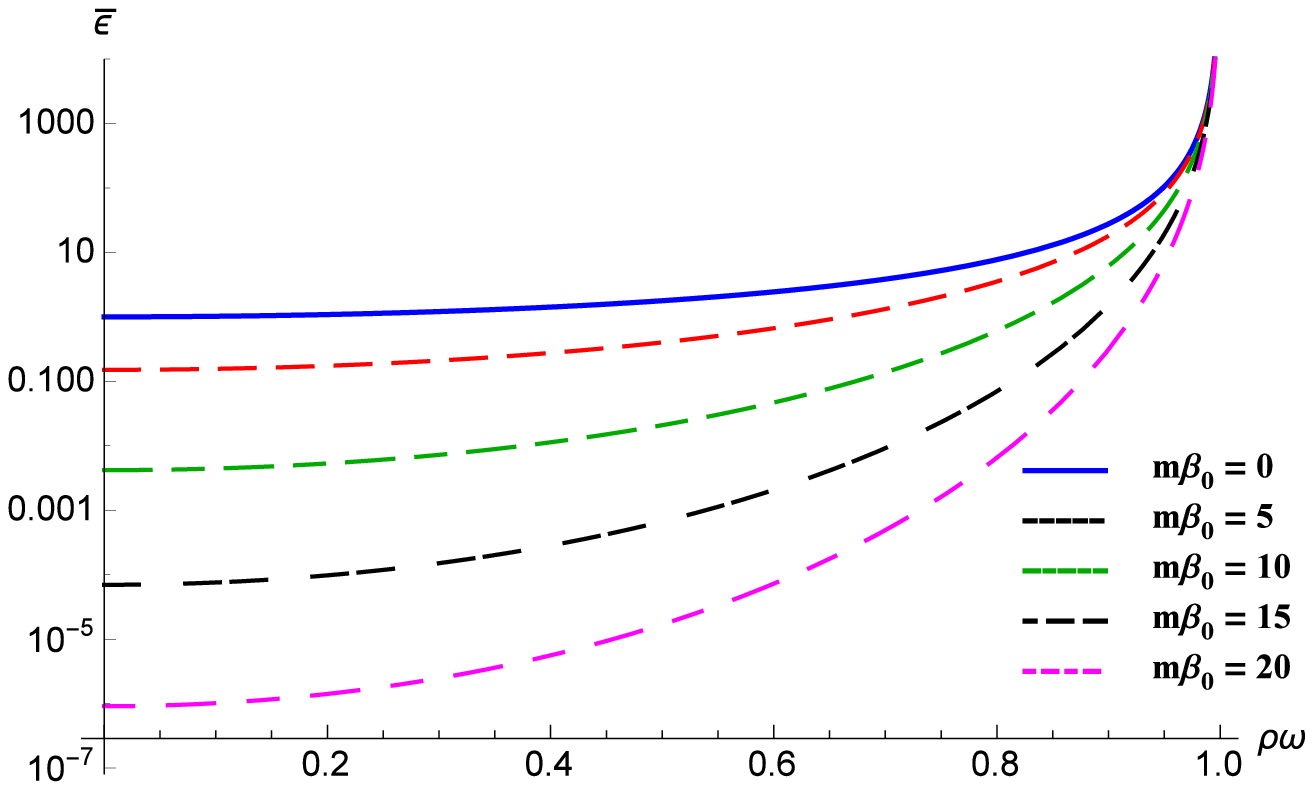} \\
(c) & (d)
\end{tabular}
\caption{
(a) Error \eqref{eq:err_def} in the evaluation of the energy density 
for Fermi-Dirac (dashed lines) and Bose-Einstein (solid lines) statistics 
by truncating the sum in Eq.~\eqref{eq:epsilon_sumj} at $j = j_{\rm max}$. 
The lower curves on the plot correspond to $m\beta = 1$, while the upper curves correspond to $m\beta = 0.01$;
(b) Departure from the M-J profile for the energy density $\epsilon$ of the B-E (top lines) and 
F-D (bottom lines) profiles as measured by $\delta \epsilon$ \eqref{eq:delta_epsilon};
(c) Equation of state $w = P / \epsilon$ for five values of the $\beta_0 m$, as a function of $\rho\omega$. 
Close to the SOL, the flow enters the ultrarelativistic regime and $w \rightarrow 1/3$.
(d) $\overline{\epsilon}$ \eqref{eq:oepsilon} for $5$ values of $\beta_0 m$, representing
the M-J energy density $\epsilon$ normalised such that $\epsilon(\rho = 0) = 1$ for massless 
particles.
}
\label{fig:plots}
\end{center}
\end{figure}

For the general case an analytical expression for the F-D and B-E equilibrium fields can not be obtained. 
Fortunately, the series in Eqs.~\eqref{eq:hydro_MJ} are fastly convergent, allowing the sum to be truncated at finite values
of $j$. To estimate the error due to the truncation of the series at $j = j_{\rm max}$, we introduce the following quantity:
\begin{align}
 {\rm err}_{\varepsilon} (j_{\rm max}) =& \left| 1 - \frac{\sum_{j = 0}^{\rm jmax} \varepsilon^j\, \epsilon_{\rm M-J}[\beta(j+1)]}
 {\sum_{j = 0}^{\rm J} \varepsilon^j\, \epsilon_{\rm M-J}[\beta(j+1)]}\right|\nonumber\\
 =& \left|1 - \frac{\sum_{j = 0}^{j_{\rm max}} \widetilde{K}_2(m \beta)
 + \frac{(m\beta)^2}{6} \widetilde{K}_1(m \beta)}{\sum_{j = 0}^J \widetilde{K}_2(m \beta)
 + \frac{(m\beta)^2}{6} \widetilde{K}_1(m \beta)} \right|,\label{eq:err_def}
\end{align}
where we set $J = 100$ and $\varepsilon = 1$ and $-1$ for B-E and F-D statistics, respectively. 
The second equality in Eq.~\eqref{eq:err_def} shows that ${\rm err}$ depends 
only on the product $m\beta$. It is thus sufficient to check the convergence for small and large values 
of $m\beta$. We have found that $j_{\rm max} = 10$ is sufficient for $0.1\%$ accuracy for arbitrarily 
small values of $m\beta$. 

To analyse the difference between the M-J, B-E and F-D statistics, we consider the deviation of the 
energy density $\epsilon_\varepsilon$ corresponding to the B-E ($\varepsilon = 1$) and
F-D ($\varepsilon = -1$) statistics with respect to M-J ($\varepsilon = 0$) statistics having the 
same number of degrees of freedom $Z$ as the B-E and F-D distributions, respectively.
For this purpose, we construct 
\begin{align}
 \delta \epsilon_\varepsilon \equiv& \frac{\epsilon_{\varepsilon}}{s_\varepsilon \epsilon_0} - 1\nonumber\\
 =& \frac{1}{s_\varepsilon} \sum_{j = 0}^\infty \frac{\varepsilon^j}{(j+1)^4} \frac{\widetilde{K}_2[\gamma^{-1} m\beta_0(j+1)]
 + \frac{1}{6}\left[\gamma^{-1} m\beta_0(j+1)\right]^2 \widetilde{K}_1[\gamma^{-1} m\beta_0(j+1)]}
 {\widetilde{K}_2(\gamma^{-1} m\beta_0)
 + \frac{1}{6}\left(\gamma^{-1} m\beta_0\right)^2 \widetilde{K}_1(\gamma^{-1} m\beta_0)},
 \label{eq:delta_epsilon}
\end{align}
where the normalisation factor $s_\varepsilon$ is introduced to account for the proportionality factors 
between the massless limits of $\epsilon_{\rm B-E}$ \eqref{eq:epsilon_BE_smallm} and $\epsilon_{\rm F-D}$ 
\eqref{eq:epsilon_FD_smallm} relative to $\epsilon_{\rm M-J}$ \eqref{eq:epsilon_MJ_smallm}. Its value is 
given by:
\begin{equation}
 s_{-1} \equiv s_{\rm F-D} = \frac{7\pi^4}{720}, \qquad 
 s_{0} \equiv s_{\rm M-J} = 1, \qquad 
 s_{1} \equiv s_{\rm B-E} = \frac{\pi^4}{90}.
 \end{equation}
It can be seen from Eq.~\eqref{eq:delta_epsilon} that $\delta \epsilon_\varepsilon$ only 
depends on the quantities $\beta_0 m$ and $\rho \omega$. As shown in Figure~\ref{fig:plots}(b), 
the energy density $\epsilon_1$ corresponding to B-E statistics becomes more energetic than the corresponding
M-J energy density $\epsilon_0$ as $\beta m$ increases (large mass or small temperatures). For 
a fixed value of $\beta_0 m$, $\delta \epsilon_1$ decreases to $0$ as the SOL is approached. 
On the contrary, in the F-D statistics, $\epsilon_{-1}$ decreases with $\beta_0 m$, while 
for fixed $\beta_0 m$, it increases with $\rho \omega$.

Further, to gain insight on the behaviour of the isotropic pressure $P$ with respect to the energy density $\epsilon$,
it is instructive to consider the equation of state $w$ \eqref{eq:w}.
Figure~\ref{fig:plots}(c) shows $w$ for the Maxwell-J\"uttner distribution in terms 
of $\rho\omega$ for five values of $\beta_0 m$. When massless particles are considered, $w = 1/3$ everywhere.
As the mass increases, $w$ decreases, in agreement with the minus sign in Eq.~\eqref{eq:w_smallmb_MJ}.
For a fixed value of $\beta m$, $w$ increases with $\rho \omega$, reaching $1/3$ as the SOL is approached.

Finally, Figure~\ref{fig:plots}(d) shows the ratio between the $M-J$ energy density $\epsilon(\rho)$ and 
the value of $\epsilon$ on the rotation axis for massless particles:
\begin{equation}\label{eq:oepsilon}
 \overline{\epsilon} = \frac{\epsilon(m,\rho)}{\epsilon(m = 0, \rho = 0)} = 
 \gamma^4 \left[\widetilde{K}_2(\gamma^{-1} \beta_0m) + \frac{1}{6}(\gamma^{-1}\beta_0 m)^2 
 \widetilde{K}_1(\gamma^{-1} \beta_0 m)\right].
\end{equation}
As before, $\overline{\epsilon}$ also depends only on $m\beta_0$ and $\rho \omega$. 
As the SOL is approached, the argument of the modified Bessel functions above tends to $0$
and the $\gamma^4$ prefactor becomes dominant, such that close to the SOL, the profiles corresponding 
to different $\beta_0 m$ overlap, diverging at the same rate.

\subsubsection{Stress-energy tensor}

Using Eqs.~\eqref{eq:L_final}, the rest frame components $T^{\mu\nu}$ of the 
stress-energy tensor (SET) can be obtained from the tetrad components $T^{\hmu\hnu}$ as follows:
\begin{equation}
 T^{\mu\nu} = L^\mu{}_\hmu L^\nu{}_\hnu T^{\hmu\hnu}=
 \begin{pmatrix}
  \gamma^2(\epsilon + p) - p & 0 & \omega \gamma^2(\epsilon + p) & 0\\
  0 & p & 0 & 0\\
  \omega \gamma^2(\epsilon + p) & 0 & \rho^{-2}\gamma^2 (\epsilon + p) - \rho^{-2} \epsilon & 0\\
  0 & 0 & 0 & p
 \end{pmatrix}.
\end{equation}
It is instructive to consider the massless limit shown in Table~\ref{tab:tmunu_m0}. 

The Bose-Einstein results coincide with the Planckian forms reported in Ref.~\cite{duffy07}.
However, quantum field theory yields an infinite thermal expectation value of the SET 
for the Klein-Gordon field throughout the space-time \cite{duffy07,ambrus14,ambrus_phd}.

In Ref.~\cite{ambrus14,ambrus_phd}, the thermal expectation value of the SET is computed analytically 
for Dirac fermions using quantum field theory. Our results recover exactly the $\beta^{-4}$ terms 
when the components of the SET are expressed with respect to the (static) tetrad, however, the 
results of Ref.~\cite{ambrus14,ambrus_phd} contain non-trivial quantum corrections of order 
$\beta^{-2}$. A more detailed analysis of these quantum corrections is postponed for future work.

\renewcommand{\arraystretch}{2}
\begin{table}
\begin{center}
\caption{
Energy density $\epsilon$, isotropic pressure $P$ and non-vanishing components of the stress-energy 
tensor $T^{\mu\nu}$ with respect to the cylindrical coordinate system \eqref{eq:ds2} 
for massless particles corresponding to the M-J, B-E and F-D statistics. The component $T^{zz}$
is equal to $T^{\rho\rho}$. The number of degrees of freedom $Z$ is $1$ for the neutral spin $0$ Maxwell-J\"uttner 
and Bose-Einstein particles, while for the F-D statistics, its value is set to $4$ to account for the 
two spins of Dirac fermions and for the antiparticles species.
}
\begin{tabular}{l|l|l|l}
 & M-J & B-E & F-D\\\hline
$\epsilon$ & 
${\displaystyle \frac{3}{\pi^2\beta_0^4} \gamma^4 }$ & 
${\displaystyle \frac{\pi^2}{30\beta_0^4}\gamma^4 }$ & 
${\displaystyle \frac{7\pi^2}{60\beta_0^4} \gamma^4 }$ \\
$P$ & 
${\displaystyle \frac{1}{\pi^2\beta_0^4} \gamma^4 }$ & 
${\displaystyle \frac{\pi^2}{90\beta_0^4}\gamma^4 }$ & 
${\displaystyle \frac{7\pi^2}{180\beta_0^4} \gamma^4 }$ \\\hline 
$T^{tt}$ & 
${\displaystyle \frac{3}{\pi^2\beta_0^4} \left(\tfrac{4}{3}\gamma^6 - \tfrac{1}{3}\gamma^4\right) }$ & 
${\displaystyle \frac{\pi^2}{30\beta_0^4} \left(\tfrac{4}{3}\gamma^6 - \tfrac{1}{3}\gamma^4\right)  }$ & 
${\displaystyle \frac{7\pi^2}{60\beta_0^4} \left(\tfrac{4}{3}\gamma^6 - \tfrac{1}{3}\gamma^4\right) }$ \\
$T^{t\varphi}$ & 
${\displaystyle \frac{4\omega}{\pi^2\beta_0^4} \gamma^6 }$ & 
${\displaystyle \frac{2\omega\pi^2}{45\beta_0^4}\gamma^6 }$ & 
${\displaystyle \frac{7\omega \pi^2}{45\beta_0^4} \gamma^6 }$ \\
$T^{\rho\rho}$ & 
${\displaystyle \frac{1}{\pi^2\beta_0^4} \gamma^4 }$ & 
${\displaystyle \frac{\pi^2}{90\beta_0^4}\gamma^4 }$ & 
${\displaystyle \frac{7\pi^2}{180\beta_0^4} \gamma^4 }$ \\
$T^{\varphi\varphi}$ & 
${\displaystyle \frac{1}{\rho^2 \pi^2\beta_0^4} \left(4\gamma^6 - 3\gamma^4\right) }$ & 
${\displaystyle \frac{\pi^2}{90\rho^2 \beta_0^4} \left(4\gamma^6 - 3\gamma^4\right)  }$ & 
${\displaystyle \frac{7\pi^2}{720\rho^2\beta_0^4} \left(4\gamma^6 - 3\gamma^4\right) }$ \\
\hline 
\end{tabular}
\label{tab:tmunu_m0}
\end{center}
\end{table}

\section{Conclusion}\label{sec:conc}

In this paper, we have considered an application of the tetrad formalism introduced in Ref.~\cite{cardall15} to the
problem of a rigidly rotating relativistic gas in Minkowski space,
which can provide a basis for the construction of lattice Boltzmann models where the momentum
space is completely decoupled from the space-time metric.
We further analyse analytically and numerically the properties of the profiles of the particle number
density $n$, energy density $\epsilon$ and isotropic pressure $P$ corresponding to the Maxwell-J\"uttner,
Bose-Einstein and Fermi-Dirac distributions and highlight their divergent behaviour as the speed of light 
surface is approached. We also report a comparison of the massless limit of the stress-energy tensor 
and results available in the literature.

\section*{Acknowledgements}

This work was supported by a grant of the Romanian National Authority for Scientific Research and Innovation,
CNCS-UEFISCDI, project number PN-II-RU-TE-2014-2910. The authors would like to express their gratitude to 
Nistor Nicolaevici for fruitful discussions. V.E.A. would also like to thank Peter Millington and Elizabeth 
Winstanley for preliminary discussions and for suggesting the comparison with the results from quantum 
field theory.

\appendix

\titleformat{\section}{\bfseries}{\appendixname~\thesection .}{0.5em}{}
\section{Momentum decomposition}\label{app:momdec}

In the previous section, the transition to spherical coordinates in the momentum space was performed.
For completeness, this section describes the momentum decomposition of the transformation matrix
$P^\ti{}_{\hati}$:
\begin{equation}
 P^\ti{}_{\hati} = \frac{1}{p} Q^\ti p_{\hati} + U^\ti{}_\hati,
\end{equation}
where
\begin{equation}
 Q^\ti = \frac{1}{p} P^\ti{}_\hati p^\hati
\end{equation}
is the projection of $P^\ti{}_\hati$ along $p^\hati$ and
\begin{equation}
 U^\ti{}_\hati = P^\ti{}_\hatj k^\hatj{}_\hati
\end{equation}
is the projection of $P^\ti{}_\hati$ orthogonal to $p^\hati$, written in terms of
the momentum space orthogonal projector:
\begin{equation}
 k^\hatj{}_\hati = \delta^\hatj{}_\hati - \frac{1}{p^2} p^\hatj p_\hati.
\end{equation}
For our example, we find:
\begin{equation}
 Q^\ti = \frac{1}{p} p^\hati \frac{\partial p^\ti}{\partial p^\hati} = \frac{\partial p^\ti}{p}
 = (1,0,0)^T,\label{eq:Q_contra}
\end{equation}
where we remind the reader that $p^\ti = (p, \theta_p, \varphi_p)^T$.
Using $P^{\tilde{1}}{}_{\hati} = \partial p / \partial p^{\hati} = p^{\hati} / p$,
the matrix $U^\ti{}_\hati$ has the following components:
\begin{equation}
 U^{\tilde{1}}{}_\hati = 0, \qquad
 U^{\tilde{2}}{}_\hati = P^{\tilde{2}}{}_\hati, \qquad
 U^{\tilde{3}}{}_\hati = P^{\tilde{3}}{}_\hati.\label{eq:U_th}
\end{equation}

Similarly, the matrix $P^\hati{}_\ti$ can also be projected:
\begin{equation}
 P^\hati{}_\ti = \frac{1}{p} p^\hati Q_{\ti} + U^\hati{}_\ti,
\end{equation}
where
\begin{equation}
 Q_\ti = P^\hati{}_\ti p_\hati = \frac{\partial p}{\partial p^\ti} = (1,0,0).
 \label{eq:Q_cov}
\end{equation}
Using $P^\hati{}_{\tilde{1}} = \partial p^\hati / \partial p = p^\hati / p$, the
components of $U^\hati{}_\ti$ are:
\begin{equation}
 U^\hati{}_{\tilde{1}} = 0, \qquad
 U^\hati{}_{\tilde{2}} = P^\hati{}_{\tilde{2}}, \qquad
 U^\hati{}_{\tilde{3}} = P^\hati{}_{\tilde{3}}.\label{eq:U_ht}
\end{equation}

Combining Eqs.~\eqref{eq:Q_contra} and \eqref{eq:Q_cov} confirms Eq.~(114) from Ref.~\cite{cardall15}:
\begin{equation}
 Q_\ti Q^\ti = 1.
\end{equation}
Similarly, Eqs.~\eqref{eq:Q_contra} and \eqref{eq:U_th} and Eqs.~\eqref{eq:Q_cov} and \eqref{eq:U_ht}
can be used to show:
\begin{equation}
 U^\hati{}_\ti Q^\ti = 0, \qquad Q_\ti U^\ti{}_\hati = 0.
\end{equation}
Finally, Eq.~(117) from Ref.~\cite{cardall15} can be checked as follows:
\begin{equation}
 U^\hati{}_\ti U^\ti{}_\hatj = P^\hati{}_{\tilde{2}} P^{\tilde{2}}{}_\hatj +
 P^\hati{}_{\tilde{3}} P^{\tilde{3}}{}_\hatj = \delta^\hati{}_\hatj -
 P^\hati{}_{\tilde{1}} P^{\tilde{1}}{}_\hatj = \delta^\hati{}_\hatj - \frac{p^\hati p^\hatj}{p^2},
\end{equation}
where the properties $P^\hati{}_\ti P^\ti{}_\hatj = \delta^\hati{}_\hatj$,
$P^\hati{}_{\tilde{1}} = p^\hati/p$ and $P^{\tilde{1}}{}_\hati = p^\hati/p$ were used.


\thebibliography{100}
\setlength{\parskip}{0pt}
\setlength{\itemsep}{3pt plus 0.3ex}
%
\bibitem{cercignani02}
C.~Cercignani, G.~M.~Kremer, {\em The relativistic {B}oltzmann equation: theory and applications},
Birkh\"{a}user Verlag, Basel, Switzerland (2002).
%
\bibitem{mendoza10prl}
M.~Mendoza, B.~Boghosian, H.~J.~Herrmann, S.~Succi, Phys.~Rev.~Lett.~{\bf 105} (2010) 014502.
%
\bibitem{lindquist66}
R.~W.~Lindquist, Ann.~Phys.~(N.~Y.) {\bf 37} (1966) 487.
%
\bibitem{riffert86}
H.~Riffert, Astrophys.~J.~{\bf 310} (1986) 729.
%
\bibitem{cardall15}
C.~Y.~Cardall, E.~Endeve, A.~Mezzacappa, Phys.~Rev.~D {\bf 88} (2015) 023011.
%
\bibitem{romatschke11}
P.~Romatschke, M.~Mendoza, S.~Succi, Phys.~Rev.~C {\bf 84} (2011) 034903.
%
\bibitem{romatschke12}
P.~Romatschke, Phys.~Rev.~D {\bf 85} (2012) 065012.
%
\bibitem{misner73}
C.~W.~Misner, K.~Thorne, J.~A.~Wheeler, {\em Gravitation},
W.~H.~Freeman and company, Oxford (1973).
%
\bibitem{nist}
F.~W.~J.~Olver, D.~W.~Lozier, R.~F.~Boisvert, C.~W.~Clark,
{\em {NIST} Handbook of Mathematical Functions},
Cambridge University Press, New York (2010).
%
\bibitem{gradshteyn}
I.~S.~Gradshteyn, I.~M.~Ryzhik, {\em Table of integrals, series and products}, 
$7$th edition, Academic Press (2007).
%
\bibitem{duffy07}
G.~Duffy and A.~C.~Ottewill, Phys.~Rev.~D {\bf 67}, 044002 (2003).
%
\bibitem{ambrus14}
V.~E.~Ambru\cb{s} and E.~Winstanley, Phys.~Lett.~B {\bf 734} (2014)  296.
%
\bibitem{ambrus_phd}
V.~E.~Ambru\cb{s}, PhD thesis, University of Sheffield (2014).

\end{document}